\documentclass[12pt,preprint,natbib,iop]{emulateapj} %Use this for all other purposes (emulateapj)
\usepackage{graphicx,epsfig,amsmath} %Comment out hyperref for submission

\lefthead{van der Wel et al.}
\righthead{Extreme Emission Line Galaxies}
\slugcomment{Version: \today}
\begin{document}

\newcommand{\kms}{\>{\rm km}\,{\rm s}^{-1}}
\newcommand{\reff}{R_{\rm{eff}}}
\newcommand{\msol}{M_{\odot}}

\title{Extreme Emission Line Galaxies in CANDELS: Broad-Band Selected,
  Star-Bursting Dwarf Galaxies at $z>1$}

%\footnote{Based on observations
%    with the Hubble Space Telescope, obtained at the Space Telescope
%    Science Institute, which is operated by AURA, Inc., under NASA
%    contract NAS 5-26555; in part on observations collected at the
%    European Southern Observatory, Chile (168.A-0485); and in part on
%    observations made with the Spitzer Space Telescope, which is
%    operated by the Jet Propulsion Laboratory, California Institute of
%    Technology, under NASA contract 1407.}

\author{A.~van der Wel\altaffilmark{1},
  A.~N.~Straughn\altaffilmark{2}, 
  H.-W.~Rix\altaffilmark{1},
  S.~L.~Finkelstein\altaffilmark{3},
  A.~M.~Koekemoer\altaffilmark{4},
  B.~J.~Weiner\altaffilmark{5},
  S.~Wuyts\altaffilmark{6}, 
  E.~F.~Bell\altaffilmark{7},
  S.~M.~Faber\altaffilmark{8},
  J.~R.~Trump\altaffilmark{8},
  D.~C.~Koo\altaffilmark{8},
  H.~C.~Ferguson\altaffilmark{4},
  C.~Scarlata\altaffilmark{9}, 
  N.~P.~Hathi\altaffilmark{10},
  J.~S.~Dunlop\altaffilmark{11},
  J.~A.~Newman\altaffilmark{12},
  M.~Dickinson\altaffilmark{13},
  K.~Jahnke\altaffilmark{1},
  B.~W.~Salmon\altaffilmark{3},
  D.~F.~de Mello\altaffilmark{14},\altaffilmark{15},
  D.~D.~Kocevski\altaffilmark{8}, 
  K.~Lai\altaffilmark{8},
  N.~A~.Grogin\altaffilmark{4},
  S.~A.~Rodney\altaffilmark{16},
  Yicheng~Guo\altaffilmark{17},
  E.~G.~McGrath\altaffilmark{8},
  K.-S.~Lee\altaffilmark{18},
  G.~Barro\altaffilmark{8},
  K.-H.~Huang\altaffilmark{16},
  A.~G.~Riess\altaffilmark{4},\altaffilmark{16}
  M.~L.~N.~Ashby\altaffilmark{19},
  S.~P.~Willner\altaffilmark{19}
}

\altaffiltext{1}{Max-Planck Institut f\"ur Astronomie, K\"onigstuhl
  17, D-69117, Heidelberg, Germany; e-mail:vdwel@mpia.de}

\altaffiltext{2}{Astrophysics Science Division, Goddard Space Flight
  Center, Code 665, Greenbelt, MD 20771, USA}

\altaffiltext{3}{George P. and Cynthia Woods Mitchell Institute for
  Fundamental Physics and Astronomy, Department of Physics \&
  Astronomy, Texas A\&M University, College Station, TX 77843, USA}

\altaffiltext{4}{Space Telescope Science Institute, 3700 San Martin
  Drive, Baltimore, MD 21218, USA}

\altaffiltext{5}{Steward Observatory, 933 N. Cherry St., University of
  Arizona, Tucson, AZ 85721, USA}

\altaffiltext{6}{Max-Planck-Institut f\"ur Extraterrestrische Physik,
  Giessenbachstrasse, D-85748 Garching, Germany}

\altaffiltext{7}{Department of Astronomy, University of Michigan, 500
  Church Street, Ann Arbor, Michigan, 48109, USA}

\altaffiltext{8}{UCO/Lick Observatory, Department of Astronomy and
  Astrophysics, University of California, Santa Cruz, CA 95064, USA}

\altaffiltext{9}{Minnesota Institute for Astrophysics, University of
  Minnesota, 116 Church St. S.E.  Minneapolis, MN 55455, USA}

\altaffiltext{10}{Observatories of the Carnegie Institution of
  Washington, Pasadena, CA 91101, USA}

\altaffiltext{11}{Institute for Astronomy, University of Edinburgh,
  Royal Observatory, Edinburgh EH9 3HJ, UK}

\altaffiltext{12}{Department of Physics and Astronomy, University of
  Pittsburgh, 3941 O'Hara Street, Pittsburgh, PA 15260, USA}

\altaffiltext{13}{National Optical Astronomy Observatory, 950 North
  Cherry Avenue, Tucson, AZ 85719, USA}

\altaffiltext{14}{Department of Physics, The Catholic University of
  America, Washington DC 20064, USA}

\altaffiltext{15}{Observational Cosmology Laboratory, Goddard Space
  Flight Center, Code 665, Greenbelt, MD 20771, USA}

\altaffiltext{16}{Department of Physics and Astronomy, Johns Hopkins
  University, Baltimore, MD 21218, USA}

\altaffiltext{17}{Astronomy Department, University of Massachusetts,
  710 N. Pleasant Street, Amherst, MA 01003, USA}

\altaffiltext{18}{Yale Center for Astronomy and Astrophysics, 260
  Whitney Avenue, JWG 454 New Haven, CT 06511, USA}

\altaffiltext{19}{Harvard-Smithsonian Center for Astrophysics, 60
  Garden Street, Cambridge, MA 02138, USA}

\begin{abstract}
  We identify an abundant population of extreme emission line galaxies
  (EELGs) at redshift $z\sim 1.7$~in the Cosmic Assembly Near-IR Deep
  Extragalactic Legacy Survey (CANDELS) imaging from \textit{Hubble
    Space Telescope}/\textit{Wide Field Camera 3} (HST/WFC3).  69 EELG
  candidates are selected by the large contribution of exceptionally
  bright emission lines to their near-infrared broad-band magnitudes.
  Supported by spectroscopic confirmation of strong [OIII] emission
  lines -- with rest-frame equivalent widths $\sim1000\rm{\AA}$ -- in
  the four candidates that have HST/WFC3 grism observations, we
  conclude that these objects are galaxies with $\sim 10^8~\msol$ in
  stellar mass, undergoing an enormous starburst phase with
  $M_*/\dot{M}_*$ of only $\sim 15$ Myr.  These bursts may cause
  outflows that are strong enough to produce cored dark matter
  profiles in low-mass galaxies.  The individual star formation rates
  and the co-moving number density ($3.7 \times 10^{-4} ~
  \rm{Mpc}^{-3}$) can produce in $\sim$4 Gyr much of the stellar mass
  density that is presently contained in $10^8-10^9~\msol$ dwarf
  galaxies.  Therefore, our observations provide a strong indication
  that many or even most of the stars in present-day dwarf galaxies
  formed in strong, short-lived bursts, mostly at $z>1$.
\end{abstract}

\section{Introduction}
The formation history of dwarf galaxies with masses $\sim 10^8~\msol$
can usually only be studied through 'archaeological' age
reconstruction, based on resolved stellar populations
\cite[e.g.,][]{grebel97, mateo98, weisz11}.  Their high-redshift
progenitors have so far remained elusive despite the ever increasing
depth of spectroscopic observing campaigns and imaging from the ground
and the \textit{Hubble Space Telescope} (HST).  In this paper we
identify an abundant population of $z>1$ dwarf galaxies undergoing
extreme starbursts, through HST/\textit{Wide Field Camera 3} (WFC3)
imaging from the Cosmic Assembly Near-IR Deep Extragalactic Legacy
Survey \citep[CANDELS,][]{grogin11, koekemoer11}, that may well be the
progenitors of present-day dwarf galaxies with stellar masses $\sim
10^8-10^9~\msol$.

At the present day, starbursts contribute a minority to the total star
formation activity in dwarf galaxies \citep{lee09}.  However, there is
abundant evidence that the star formation histories are complex and
that bursts play an important role \citep[as reviewed
by][]{mateo98}. Many authors find evidence for short-lived ($\sim$10
Myr) SF events in nearby star-forming dwarf galaxies from a range of
observational and modeling techniques \citep[e.g.,][]{schaerer99,
  mas99, thornley00, tremonti01, harris04}, while others argue that
star formation epochs of dwarf galaxies are more prolonged
\citep[e.g.,][]{calzetti97, lee08, mcquinn09}.  Simulations also
indicate that star formation histories of low-mass galaxies are
episodic or even burst-like \citep[e.g.,][]{pelupessy04, stinson07,
  nagamine10}.

As most stars in dwarf galaxies formed more than 5 Gyr ago
\citep[e.g.,][]{dolphin05, weisz11}, it is crucial to understand the
mode of star formation in dwarf galaxies at those early epochs, but
'archaeological' studies do not have the resolution in terms of
stellar population age to constrain strengths, durations, and
frequency of bursts.  The increased frequency of interactions with
other galaxies and higher gas fractions at $z>1$ may have resulted in
strong, short-lived starbursts.  In this paper we place the first
constraints on the open question of how many and how frequently
strong, short-lived starbursts occur in dwarf galaxies at $z>1$, and
how relevant this mode of star formation is for the build-up of the
dwarf galaxy population in a cosmological context.

\begin{figure}[t]
\epsscale{1.2} 
%\epsscale{0.5} 
\plotone{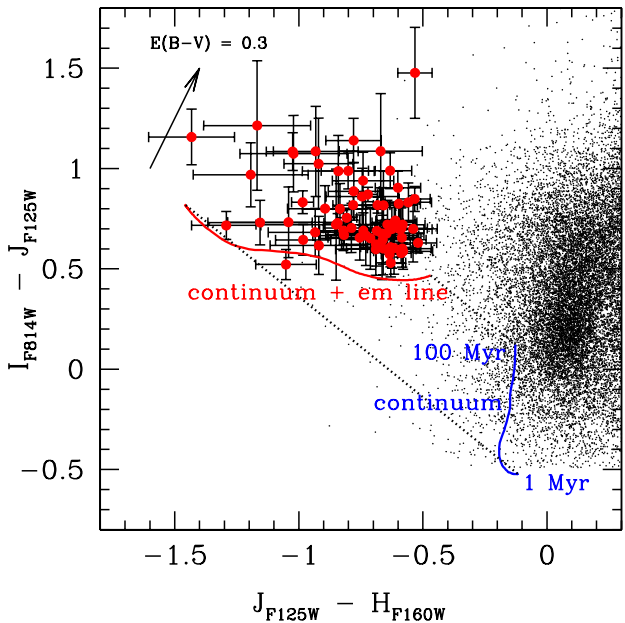}
%\plotone{seds.ps}
%\plotone{spec.ps}
\caption{ Observed $I-J$ vs.~$J-H$ colors (in AB magnitudes) from
  HST/WFC3 and ACS imaging for all objects in the UDS and GSD (small
  black points) and the sample of emission-line dominated objects
  (large red points with error bars), selected by
  $I-J>0.44+\sigma(I-J)$ and $J-H<-0.44-\sigma(J-H)$, where
  $\sigma(I-J)$ and $\sigma(J-H)$ are the $1\sigma$ uncertainties on
  the colors.  The blue line represents the redshifted ($z = 1.7$)
  continuum colors of the Starburst99 model \citep{leitherer99} for
  continuous star formation, in the age range from 1 Myr to 100 Myr.
  The red line represents the same model, but with the $J$-band flux
  density increased by the emission line luminosity predicted by the
  model (Starburst99 predicts H$\alpha$ luminosity -- [OIII] emission,
  which falls in the $J$ band at $z=1.7$, is assumed to have the same
  equivalent width). The black arrow indicates dust attenuation.}
\label{col}
\end{figure}

\section{Data}

\subsection{Multi-Wavelength Imaging}

We select objects from multi-wavelength photometry of two fields with
HST/WFC3 and \textit{Advanced Camera for Surveys} (ACS) coverage: the
Ultra Deep Survey (UDS) field and the GOODS-South Deep (GSD) field at
4-epoch depth.\footnote{GOODS is the Great Observatories Origins Deep
  Survey. CANDELS provides deep images over the central parts of
  GOODS-North and GOODS-South, and wider, less deep imaging over the
  remainder of those fields and over the other CANDELS fields,
  including UDS.  See the CANDELS website, http://candels.ucolick.org/
  , for details of the field layouts.} For the UDS we use WFC3 imaging
in F125W ($J$) and F160W ($H$) and ACS imaging in F814W ($I$) from
CANDELS.  For the GSD we use the $J$ and $H$ band imaging from
CANDELS, supplemented by WFC3 imaging from the Early Release Science
(ERS) program \citep{windhorst11}, and $I$ band imaging from GOODS
\citep{giavalisco04}.  The total area with $I$-, $J$-, and $H$-band
coverage used here is 279 square arcminutes.

Sources are detected in the $H$ band with SExtractor \citep{bertin96}
and photometry is performed with TFIT \citep{laidler07}, which uses
additional imaging data sets, ranging from $U$ to 4.5$\mu$m to produce
resolution-matched, multi-wavelength catalogs. The catalog
construction is described in full by Guo et al.~(in prep.).  In
addition, we use a version of GALAPAGOS \citep{haussler07} adapted for
CANDELS WFC3 imaging to measure structural parameters (van der Wel et
al., in prep.).

\subsection{Color-Color Selection}

We select objects that are red in $I-J$ and blue in $J-H$ (see Figure
\ref{col}), tracing luminous emission lines that contribute
significantly to the total $J$-band light.  No known continuum
emission can produce such broad-band colors.  The highlighted objects
in Figure \ref{col} have $I-J > 0.44 + \sigma (I-J)$ and $J-H < - 0.44
- \sigma (J-H)$, where $\sigma$ refers to the color uncertainty; that
is, we select those objects that are significantly more than 50\%
brighter in J than in both I and H.  We identify 69 such objects, that
is, there is 1 per $\sim$4 square arcminutes.  They range in magnitude
from $H_{\rm{AB}}=24$ to $H_{\rm{AB}}=27$, with a median of $H=25.8$
(see Table 1).  We note that there is no gap in color-color space
between the emission line galaxy candidates that we select and the
general distribution; the selected objects are merely the most extreme
outliers.

\begin{figure*}[t]
\epsscale{1.17} 
\plotone{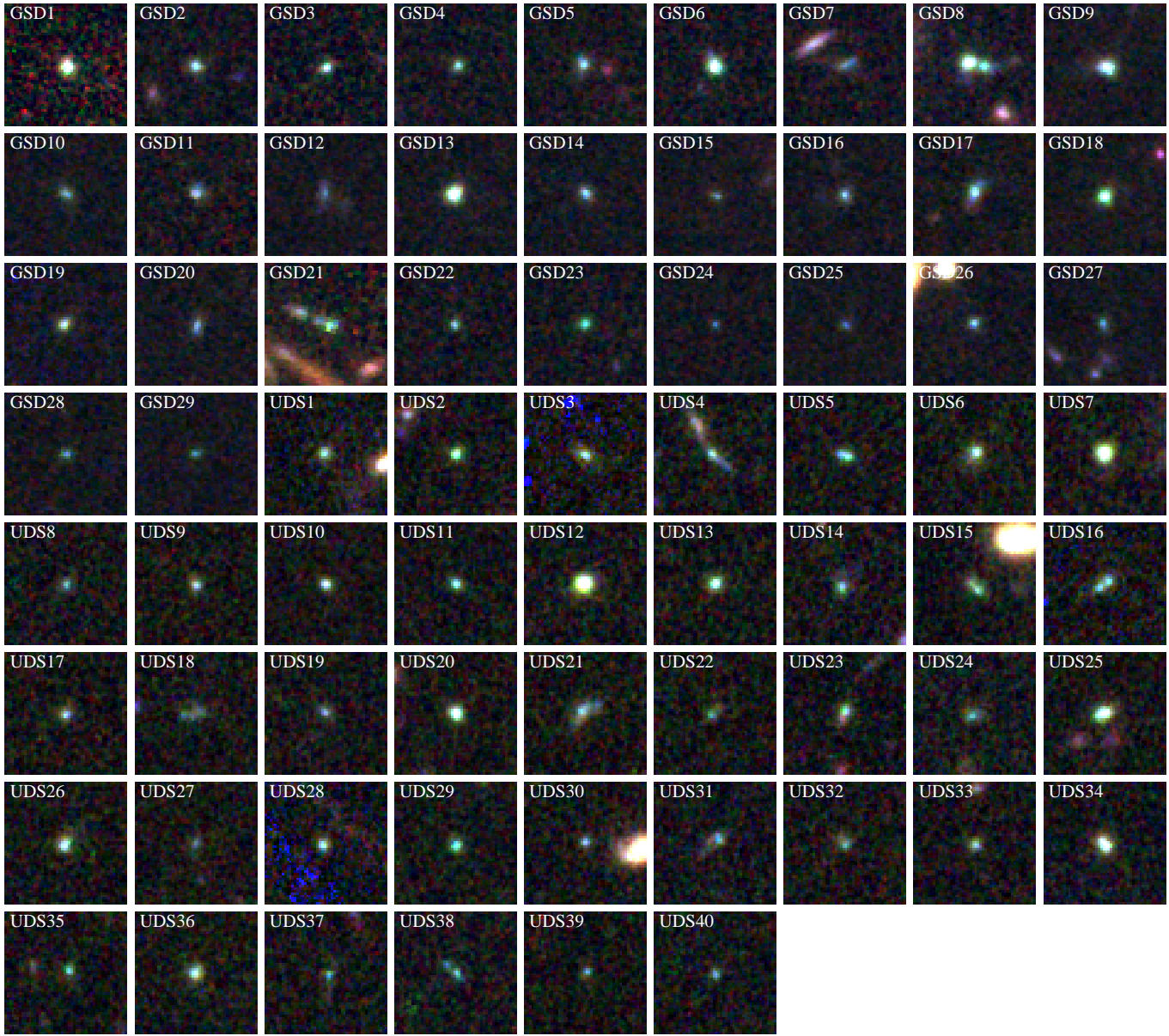}
\caption{ False-color composites, created from HST $I$, $J$, and $H$
  band image cutouts of the 69 emission line galaxy candidates.  The
  cutouts are $3\arcsec$~on a side, the pixel scale is 0.06$\arcsec$,
  and the full-width half maximum resolution at the longest wavelength
  (the $H$ band) is $\sim0.18\arcsec$. The IDs correspond to those in
  Table 1.  The sources are typically compact, although a subset of
  about 20\% have more extended morphologies or feature multiple
  components.}
\label{cutouts}
\end{figure*}

\begin{figure}[h]
\epsscale{1.2} 
\plotone{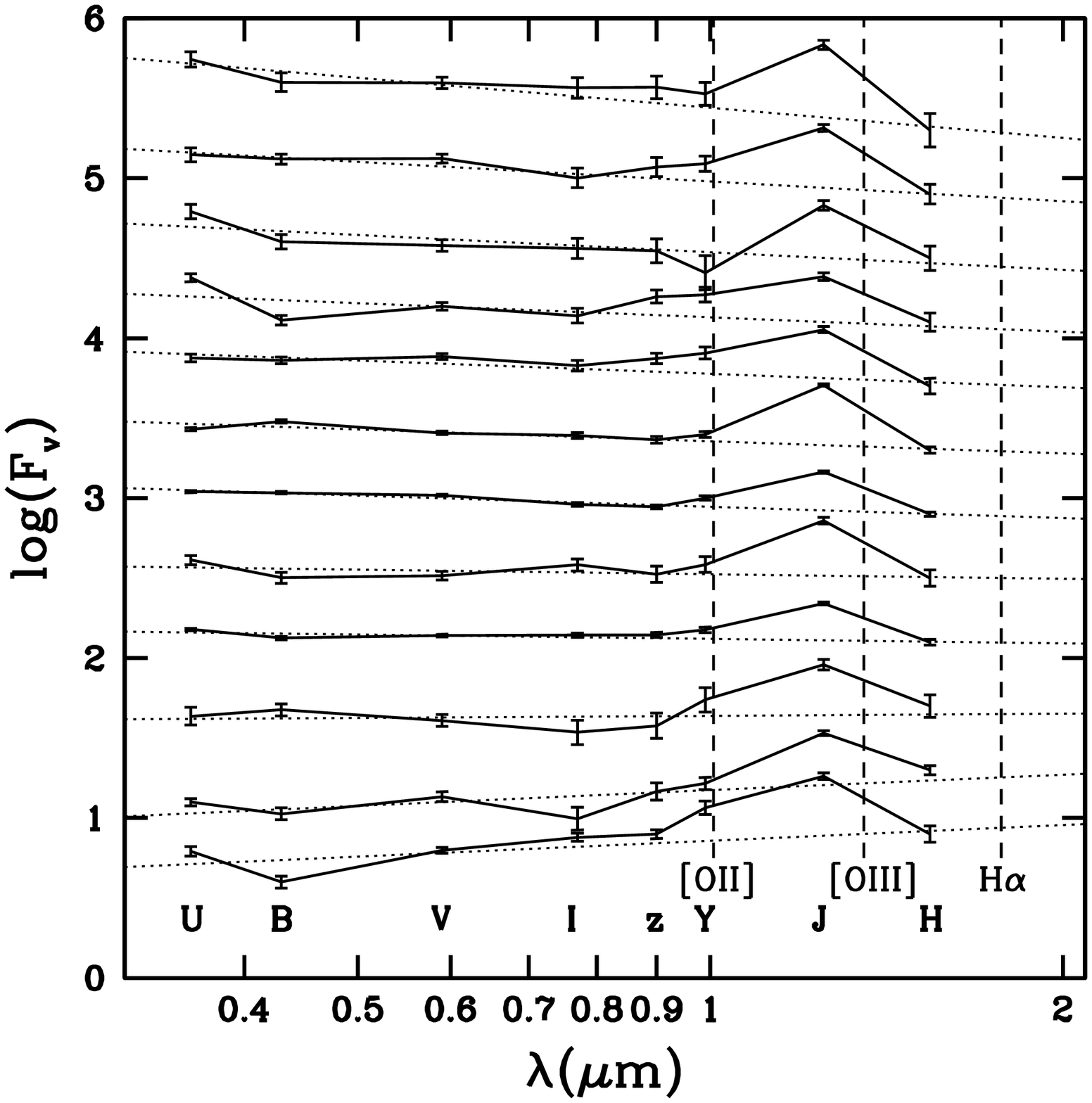}
\caption{ Broad-band SEDs of the 12 emission line galaxy candidates
  selected from the GSD field.  Units on the y-axis are arbitrary, and
  the SEDs are incrementally offset by 0.4 dex in the vertical
  direction for clarity, sorted by continuum slope, indicated by the
  dotted lines.  The objects are characterized by flat SEDs in
  $F_{\nu}$ over the entire range from $U$ band to $H$ band.  The $J$
  band noticeably deviates from this trend as the result of strong
  emission line contributions, mostly [OIII] at $z\sim 1.7$. The
  observed wavelengths for $z=1.7$ of various emission lines are
  indicated by the vertical dashed lines.}
\label{seds}
\end{figure}

In Figure \ref{cutouts} we show false-color composites of all 69
candidates.  These sources are typically compact, but not unresolved;
their $J$- and $H$-band half-light radii from GALFIT are typically
0.1$\arcsec$. A subset ($\sim 20\%$) are more extended or consist of
multiple components.  We show the $U$ through 4.5$\mu m$ spectral
energy distributions (SEDs) of a subset of the emission-line
candidates in Figure \ref{seds}.  The SEDs are seen to be almost
entirely flat in $F_{\nu}$, or in terms of ultra-violet spectral slope
they have $\beta\sim -2$, where $\beta$ is defined as
$F_{\lambda}=\lambda^{\beta}$.  The $J$ band is a notable outlier from
this SED shape for all these objects.

\section{Extremely Bright Emission Lines}

\subsection{Photometric Constraints}

No known objects have continuum SEDs that resemble those shown in
Figure \ref{seds}; in particular, the extraordinarily blue $J-H$
colors are difficult to explain by any radiative process.  Our
hypothesis is that the $J$-band excess is due to one or more emission
lines.  The implied equivalent widths in the observed frame are
extraordinarily high: $\rm{EW}\sim 1500-3000\rm{\AA}$.

Among the emission lines that can reach such extreme EWs, Ly$\alpha$
and [OII] are immediately ruled out because the implied high redshift
would produce a strong break in the SEDs; the lack of such a break
implies $z<2$ for these objects.  WFC3/UVIS observations
\citep{windhorst11} provide UV photometry over the ERS area.  The
average color of those candidates in the ERS area is $F275W - U =
1.44$, which suggests that the Lyman break is situated at around
3000$\AA$ in the observed frame, which, in in combination with the
very blue continuum slopes redward of the U band, implies $z>1.5$.
Only one candidate with UVIS coverage has a $F275W - U$ color
consistent with that of a galaxy at $z<1.5$.  The implication is that
strong [OIII] emission at 4959$\rm{\AA}$ and 5007$\rm{\AA}$ provide
the most plausible explanation for the J band excess light.

If [OIII] is responsible for the J band excess the redshift upper
limit is $z=1.8$.  Furthermore, because we select objects with blue
$J-H$ colors, the $H$ cannot contain the bright H$\alpha$ line, which
implies $z>1.6$.  Thus, solely based on their photometric properties,
we suggest that our candidates are strong [OIII] emitters in the
redshift range $1.6 < z < 1.8$.

\subsection{Spectroscopic Constraints}

\begin{figure}[t]
\epsscale{1.2} 
\plotone{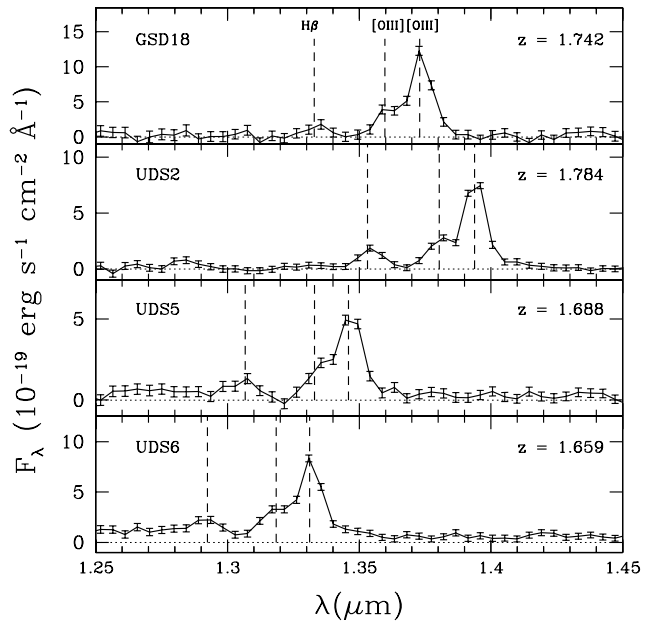}
\caption{ WFC3 grism spectra of the four candidates with grism
  coverage. The IDs refer to those in Table 1.  GSD18 is object 402
  from \citet{straughn11}; the 3 objects in the UDS are from supernova
  follow-up grism observations (program ID 12099, PI A.~Riess).  The
  three vertical dashed lines show positions of H$\beta$, [OIII], and
  H$\alpha$ for $z=1.7$.  These spectra strongly suggest that the
  majority of the objects in our sample are [OIII] emitters at $z\sim
  1.7$.}
\label{spec}
\end{figure}

The hypothesis that [OIII] emission at $z\sim 1.7$ explains the J band
excess light is strongly supported by spectroscopic observations.
While none of the candidates have ground-based spectra, WFC3 grism
observations are available for small portions of the GSD \citep[one
pointing in the ERS field,][]{straughn11} and the UDS (from the
supernova follow-up program 12099, PI A.~Riess).  The available grism
coverage overlaps with the positions of 4 candidates in our sample (1
in the ERS, 3 in the UDS), and strong emission lines are detected in
all 4 cases.  The spectra (Figure \ref{spec}) all show bright emission
lines in the wavelength range $1.3-1.4\mu m$, whose combined fluxes
are in agreement with the excess light seen in the J band.

The lines in all 4 spectra are readily identified as [OIII]: the
asymmetry of the bright line, always extended blueward, is due to the
two components of the [OIII] line, at 5007$\rm{\AA}$ and at
4959$\rm{\AA}$, where the latter is $3\times$ fainter.  In all cases,
H$\beta$ is also detected. The redshifts are all in the range
$z=1.65-1.80$, in excellent agreement with what we inferred solely
from photometry.  We conclude that our sample of extreme emission line
galaxies (EELGs) form the high-EW tail of the general population of
emission line galaxies seen in ACS and WFC3 spectroscopic grism
observations \citep[e.g.,][]{straughn08, straughn09, atek10,
  straughn11}.

In principle, H$\alpha$ emitters at $0.9<z<1.1$ should also be
included by our selection technique.  The spectroscopy and UV
photometry indicate that those must be far fewer than [OIII] emitters.
Whether this is due to selection effects or evolution in the number
density of such objects remains to be seen.  Still, even though our
working hypothesis is that all 69 candidates are [OIII] emitters at
$z\sim 1.7$, we should keep in mind that some fraction of our 69
candidates are likely H$\alpha$ emitters at $z\sim 1$.

\subsection{Emission Lines and Broad-Band Photometry}

% The existence of such strong [OIII] emitters has been demonstrated
% before.  At low redshifts, $z<0.4$, \citet{cardamone09} identified a
% rare class of emission-line dominated objects through broad-band
% colors.  Narrow-band surveys identified galaxies with strong [OIII]
% and H$\alpha$ emission lines with $\rm{EW}\sim 100-1000\rm{\AA}$ at
% redshifts $z=0.3-1$ \citep[e.g.,][]{kakazu07}, demonstrated to be
% young and metal poor \citep{hu09}.  Most notably, \citet{atek10}
% identified galaxies with [OIII] emission lines with $\rm{EW}>
% 1000\rm{\AA}$ at $z=1-1.5$; these objects are likely of the same
% nature as those in our broad-band selected sample.

We have shown that selecting objects which are much brighter in $J$
than in $I$ and $H$ works as a rather clean method for finding strong
[OIII] emitters at $1.6<z<1.8$.  Emission line galaxies with such
excesses in other bands also exist, but a systematic search is more
complicated as at most redshift ranges, multiple lines (most notably
[OIII] and H$\alpha$) affect multiple photometric bands.  Therefore,
we refrain from conducting such a systematic search here.

The existence of such emission-line dominated galaxies complicates the
interpretation of SEDs, which is especially relevant in the case of
the search for and SED modeling of rare, high-redshift objects.
Although contamination by emission lines is often considered to be a
factor \citep[e.g.,][]{labbe10}, the extremely bright lines we observe
suggests that their effect may be underestimated.  \citet{ono10}
explicitly showed that red colors in Ly$\alpha$ emitters and $z=7$
Lyman break galaxies may indicate the presence of evolved stellar
populations or strong nebular emission lines \citep[also
see][]{schaerer99, finkelstein11a}.  Steep UV continuum slopes, such
as observed in our objects, should serve as a warning sign for
contamination by nebular emission lines at longer wavelengths to the
point that those can dominate the broad-band flux density.

%to next section
%In the following we will assume that all objects are strong [OIII]
%emitters in the redshift range $1.6<z<1.8$.  We compute the equivalent
%width as follows: \begin{equation} \log{EW_{[OIII]}} = 0.4 \times
%  ((I-J)/2 - (J-H)/2)~+~\log{(W_{F125W}/(1+z))},\end{equation} where
%$W_{F125W}=2845\rm{\AA}$ is the effective width of the $J$-band
%response curve, and $z=1.7$ to correct the observed EW to the rest
%frame.  Thus, we attribute the excess light in the $J$ band, as
%compared to the $I$ and $H$ bands, to the [OIII] line contribution.

\begin{figure}[t]
\epsscale{1.2} 
\plotone{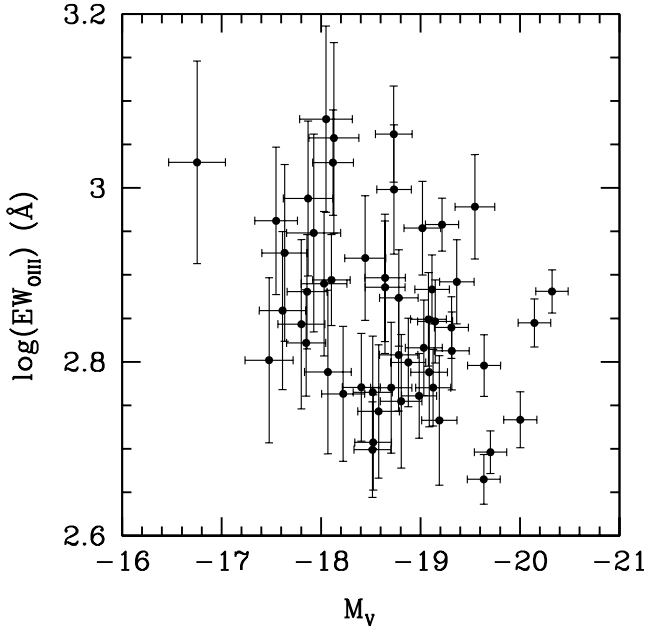}
\caption{ Rest-frame V-band absolute magnitude vs.~rest-frame
  equivalent width of the [OIII] emission line at 5007$\rm{\AA}$ as
  calculated from the broad-band photometry as explained in the text
  and assuming that all emission-line galaxy candidates are at
  $z=1.7$. They span a range in luminosity, $M_v=-17$ to $M_V=-20$,
  and have $EW_{[OIII],5007}$ between 500$\rm{\AA}$~and
  1200$\rm{\AA}$. See also Table 1.}
\label{mv_ew}
\end{figure}

\section{Starbursting Dwarf Galaxies at $z=1.7$}

\subsection{Star formation or AGN?}

Before turning to our preferred starburst interpretation, let us first
point out that nuclear activity is not a likely explanation for the
bright emission lines in the vast majority of EELGs.  None of the
objects in the CDFS have significant detections in X-ray or at
24$\mu$m. The objects are spatially resolved in both J and H, and,
moreover, the $J$ and $H$ band sizes are consistent with each other.

Moreover, it is highly unlikely that all 69 objects are dominated by
line emission from active galactic nuclei.  At least at the present
day, low-mass, low-metallicity AGN are exceedingly rare
\citep{izotov08}, much rarer than starbursting dwarf galaxies
\citep{izotov11}.  The implied black hole masses for the objects in
our sample, as inferred from their UV continuum luminosities
\citep{shen08} are $\sim 10^6~\msol$ at most, when assuming an
Eddington accretion of unity.  At these low masses, at least at the
present day, secular processes drive nuclear activity; thus, an
unknown accretion mode or triggering mechanism for nuclear activity
would have to be invoked in order to explain an extreme change in the
relative numbers of AGN and starburst powered emission-line dominated
objects.  At these low masses, merging cannot account for this.  The
starburst hypothesis, on the other hand, places these objects in the
realm of dwarf galaxies, and their physical and statistical properties
are consistent with the abundances and masses of dwarf galaxies as we
will discuss below.
%Moreover, the implied black hole masses, as inferred from their
%continuum luminosities \citep{shen08} and assuming an Eddington
%accretion rate of 1, are $\sim 10^6~\msol$.  The stellar masses
%derived below are $\sim 10^8~\msol$ -- and these turn into upper
%limits if AGN are invoked here -- which make the black hole to stellar
%mass ratio much higher than 0.001, the canonical value observed today.
%Evolution in this number by an order of magnitude is not expected.

Although nuclear activity cannot be ruled out entirely -- and
line-strength gradients in star forming $z\sim 2$ galaxies suggest
that weak AGN may contribute to some extent \citep{trump11} -- we can
safely assume that the observed emission lines are effectively
dominated by star formation activity.

\subsection{Starburst Ages and Masses}

We interpret the observations in the context of the Starburst99 model
\citep[SB99,][]{leitherer99}, which includes predictions for how the
EWs of Hydrogen recombination line evolve over time.  Therefore, our
first task is to estimate H$\beta$ line strengths from the data.  We
attribute the excess light in the J band, compared to the continuum
light measured in the I and H bands, to combined effect of emission
lines.  Therefore, we can compute the combined equivalent width as
follows:
\begin{equation} \rm{EW} = \Big(f_J-\frac{f_I+f_H}{2}\Big)
  \frac{\rm{W_J}}{1+z} \end{equation} where $W_{J}=2845\rm{\AA}$ is
the effective width of the $J$-filter response curve, $z=1.7$ to
correct the observed EW to the rest frame, and $f$ is the flux density
$f_\nu$ in the respective filters.

The relative contributions of the various emission lines are
constrained by fitting Gaussian components to the 3 emission lines
seen in the grism spectra shown in Figure \ref{spec}, keeping the
ratio between the two [OIII] components fixed at 3.  We only use the 3
UDS spectra as H$\beta$ is only marginally detected in the GSD
spectrum.  The emission line ratios are remarkably similar for all 3
objects: H$\beta$ contributes 1/8 to the combined line luminosity,
suggesting a very low metallicity \citep[see, e.g.,][for
comparisons]{salzer05, amorin10}.  Because the flux is dominated by
the [OIII]$_{5007}$ line and is therefore more directly related to our
observations, we show the inferred [OIII]$_{5007}$ EWs in Figure
\ref{mv_ew} (also see Table \ref{tab}).  However, we model the
observations by fitting the inferred H$\beta$ EWs to the SB99
predictions.  These are assumed to be always 1/8th of the combined EW.
The unavoidable intrinsic scatter in this conversion is mimicked by
propagating a generous factor of two in the uncertainties of the
quantities we infer below.

%\pontzon

\begin{figure}[t]
\epsscale{1.2} 
\plotone{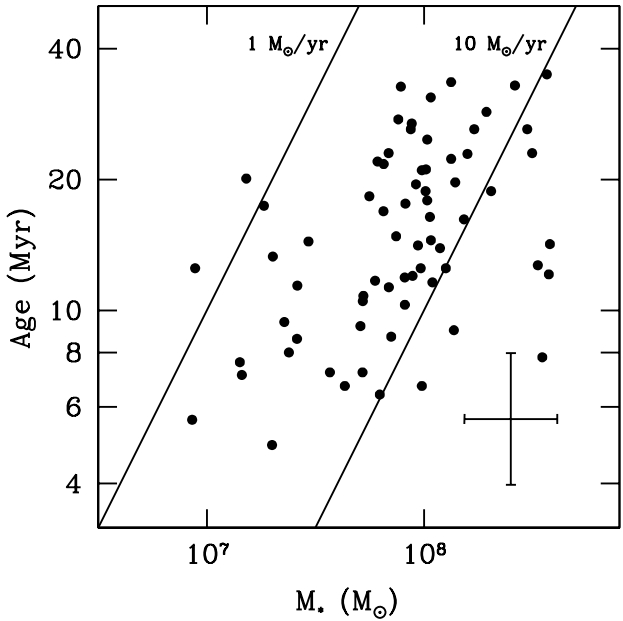}
\caption{ Masses, ages and star formation rates for the 69
  emission-line dominated objects in our sample, derived with the SB99
  model, assuming that all are at $z=1.7$, adopting a continuous SF
  model with 0.2 times solar metallicity and a Chabrier IMF.  The SFRs
  indicated by the diagonal lines are simply obtained by dividing the
  mass (x-axis) by the age (y-axis).  The galaxies in our sample
  typically have $10^8~\msol$ stellar masses with young ages ($5-30$
  Myr), or, equivalently, extremely high specific SFRs ($\sim
  5\times10^{-8}~\rm{yr}^{-1}$, or $\sim 50\times
  t_{\rm{Hubble}}^{-1}$). }
\label{m_age}
\end{figure}

$EW_{\rm{H}\beta}$ is a sensitive age indicator, as it is quickly
reduced once a stellar population gains in mass or the star formation
activity diminishes.  For a SB99 model with continuous star formation
with a \citet{chabrier03} IMF with a high-mass cut off at 100$\msol$
and metallicity 0.2$Z_{\odot}$ the H$\beta$ EWs imply that the
galaxies in our sample typically have ages of $10-20$~Myr (Figure
\ref{m_age}).  It seems unlikely that these bursts will be much longer
than this given the energy put into the interstellar medium (see
Sec.~4.3); follow-up grism observations will directly constrain the
number of older bursts.

If we assume a single burst model instead, we infer ages $3-5$~Myr;
all formation histories with declining star formation rates produce
ages that are bracketed by these two extremes.  Continuous star
formation seems more realistic than an instantaneous burst, a notion
that becomes physically untenable at very young ages.  Star formation
cannot happen faster than the dynamical time scale of $\sim$10 Myr for
these systems.  Most relevant for our analysis is that the stellar
mass estimates derived from the two different models are very similar
(see below).

We note that choosing a different metallicity does not significantly
change our results -- a low metallicity is realistic for these
low-mass systems \citep{amorin10}, and the [OIII]-H$\beta$ ratio
suggests the metallicity is indeed low.  Choosing an even lower
metallicity (0.05$Z_{\odot}$ instead of 0.2$Z_{\odot}$) results in an
increase in both age and mass by $\sim 0.2$ dex.  We do suffer from
the usual, unknown uncertainty due to our lack of knowledge of the
stellar IMF.  A different slope or cutoff at the high-mass end changes
the number of ionizing photons from high-mass stars per unit stellar
mass, and the entirely unconstrained number of low-mass stars
($\lesssim 1~\msol$) determines the overall normalization of the
stellar mass.  In general, we conclude that the ages of these galaxies
are $\lesssim40$~Myr, which includes the intrinsic range in age and
the systematic uncertainty due to the unknown star formation history.

In the following we use the results from the continuous star formation
model, but using the instantaneous burst model, by virtue of the
insensitivity of the mass estimates to the choice of star formation
history, does not change our interpretation and conclusions.

For a given age, the SB99 model predicts the (rest-frame) $V$-band
mass-to-light ratio, such that we can directly estimate the mass after
deriving the $V$-band luminosity from the observed $H$ band magnitude.
We correct the luminosity and the derived mass estimate for extinction
by comparing the continuum slope derived from the ACS photometry at
rest-frame 2500$\rm{\AA}$, typically $\beta_{2500}\sim -2$ (see Table
1), with the SB99 model prediction (rather constant at
$\beta_{2500}\sim -2.6$ for the ages of these bursts).  If we adopt
the \citet{calzetti00} extinction law for starbursting galaxies the
typical extinction is $E(B-V)=0.2$.

The median mass we infer is $8\times10^{7}~\msol$ (see Table 1 and
Figure \ref{m_age}).  Mass estimates inferred from the instantaneous
burst model are only slightly smaller, by less than 0.1 dex on
average.  Internal consistency lends our modeling approach strong
credibility: given the inferred ages, masses, and extinction
corrections, the SB99 model predicts dust-attenuated rest-frame UV
luminosities that are consistent with the observed rest-frame UV
luminosities -- the latter are not used in our modeling procedure.
Thus, the model successfully describes the observed rest-frame UV and
optical continuum spectral energy distributions as well as the
observed emission line luminosities.

Full modeling of the spectral energy distributions that includes
emission line contributions will be presented in forthcoming studies
that will also include objects with less prominent emission lines.  As
a consistency check with the results presented above we already
applied the method outlined by \citet{finkelstein11a} to the galaxies
in our sample.  This method compares the observed photometry with
model spectral energy distributions that include contributions from
nebular and stellar continuum radiation as well as all nebular
emission lines.  Free parameters in this modeling procedure include
redshift, stellar mass, extinction and metallicity.  We find
photometric redshifts that are consistent with $1.6<z<1.8$ for the
vast majority of the sample.  Moreover, the inferred stellar masses
and ages are very similar, even though this model is based on a
different stellar population synthesis model.

The implied star-formation rates (see Figure \ref{m_age}) may lead one
to expect that these galaxies have significant 24$\mu$m detections.
However, we have verified that none of the 29 candidates in the GSD,
which has very deep 24$\mu$m from GOODS, is detected.  A possible
explanation for this is the presumably low metallicity, which would
result is relatively small dust masses and, hence, low infrared
luminosities.

The observed IRAC flux densities at 3.6$\mu\rm{m}$ and 4.5$\mu\rm{m}$
from GOODS and SEDS\footnote{GOODS and SEDS (Spitzer Extended Deep
  Survey) provide the deepest IRAC imaging ever obtained, an excellent
  probe of stellar populations at high redshift} are in most cases --
there are two exceptions -- fully consistent with the expected flux
densities for the bursts observed in the UV.  In addition, the
galaxies have the same sizes in the J and H bands, indicating that the
spatial extent of the region from which the line emission originates
roughly follows the stellar light.  Hence, there is no evidence for
underlying older stellar populations.  However, we cannot rule out
their existence: maximally old stellar populations have mass-to-light
ratios that are up to $\sim50$ times larger than those of the bursts,
even in the near infrared.

If we assume a past star formation rate that is constant after
averaging over $>100$~Myr time scales we find upper limits for the
mass in older stars that is $\sim 5\times$ the burst mass.  The
implied total stellar mass upper limits are then $\lesssim 5\times
10^8~\msol$.  This caveat notwithstanding, we assume in the remainder
of this paper that there is no significant population of older stars
in these galaxies, and that the observed bursts account for the total
stellar mass.  However, the bottom line is that the total stellar
masses of these objects are well below $10^9~\msol$, in the regime of
dwarf galaxies.

\begin{deluxetable*}{lcccccc}
\tabletypesize{\scriptsize}
\tablecaption{Sample of Extreme Emission Line Galaxies\label{tab}}
\tablewidth{0pt}
\tablehead{
\colhead{EELG2011} & \colhead{RA (J2000)} & \colhead{DEC (J2000)} & \colhead{H} & \colhead{$EW_{\rm{[OIII]},5007}$} & \colhead{$\beta_{2500}$} & \colhead{$\log{(M)}$} \\
             &   (deg)   & (deg)      &     (AB)  &         ($\rm{\AA}$)            &                   &     ($\msol$)       \\
}
\startdata
GSD1 & 53.167064 & -27.858936 &      24.67 $ \pm $       0.07 &        459 $ \pm $         40 &      -1.83 $ \pm $       0.09 &       8.57 $ \pm $       0.17 \\ 
GSD2 & 53.080345 & -27.850572 &      25.63 $ \pm $       0.07 &        569 $ \pm $         67 &      -2.04 $ \pm $       0.19 &       7.96 $ \pm $       0.19 \\ 
GSD3 & 53.046020 & -27.837322 &      25.88 $ \pm $       0.11 &        507 $ \pm $         75 &      -2.02 $ \pm $       0.18 &       7.94 $ \pm $       0.20 \\ 
GSD4 & 53.105087 & -27.819974 &      26.20 $ \pm $       0.11 &        769 $ \pm $        143 &      -1.75 $ \pm $       0.30 &       7.72 $ \pm $       0.22 \\ 
GSD5 & 53.067508 & -27.773595 &      25.05 $ \pm $       0.08 &        566 $ \pm $         74 &      -2.19 $ \pm $       0.19 &       8.14 $ \pm $       0.19 \\ 
GSD6 & 53.097499 & -27.763919 &      24.99 $ \pm $       0.05 &        700 $ \pm $         53 &      -2.12 $ \pm $       0.09 &       8.10 $ \pm $       0.17 \\ 
GSD7 & 53.122127 & -27.759542 &      25.44 $ \pm $       0.12 &        535 $ \pm $         99 &      -1.70 $ \pm $       0.28 &       8.20 $ \pm $       0.22 \\ 
GSD8 & 53.171936 & -27.759145 &      24.26 $ \pm $       0.04 &        693 $ \pm $         47 &      -1.76 $ \pm $       0.11 &       8.52 $ \pm $       0.17 \\ 
GSD9 & 53.078754 & -27.750288 &      24.86 $ \pm $       0.04 &        468 $ \pm $         32 &      -1.99 $ \pm $       0.10 &       8.42 $ \pm $       0.17 \\ 
GSD10 & 53.063690 & -27.745853 &      26.37 $ \pm $       0.09 &        759 $ \pm $        134 &      -1.56 $ \pm $       0.33 &       7.72 $ \pm $       0.22 \\ 
GSD11 & 53.007499 & -27.741867 &      25.97 $ \pm $       0.09 &        534 $ \pm $         76 &      -2.13 $ \pm $       0.19 &       7.84 $ \pm $       0.20 \\ 
GSD12 & 53.114612 & -27.721979 &      25.80 $ \pm $       0.12 &        641 $ \pm $        139 &      -1.94 $ \pm $       0.31 &       7.87 $ \pm $       0.23 \\ 
GSD13 & 53.101516 & -27.720882 &      24.77 $ \pm $       0.03 &        490 $ \pm $         29 &      -2.35 $ \pm $       0.08 &       8.29 $ \pm $       0.16 \\ 
GSD14 & 53.055908 & -27.718803 &      26.00 $ \pm $       0.08 &        501 $ \pm $         65 &      -1.90 $ \pm $       0.20 &       7.94 $ \pm $       0.19 \\ 
GSD15 & 53.149536 & -27.710285 &      26.64 $ \pm $       0.15 &        820 $ \pm $        288 &      -2.20 $ \pm $       0.57 &       7.36 $ \pm $       0.30 \\ 
GSD16 & 53.147617 & -27.707088 &      26.10 $ \pm $       0.08 &        582 $ \pm $         80 &      -2.08 $ \pm $       0.22 &       7.75 $ \pm $       0.20 \\ 
GSD17 & 53.064220 & -27.706523 &      25.41 $ \pm $       0.05 &        465 $ \pm $         43 &      -2.20 $ \pm $       0.13 &       8.12 $ \pm $       0.17 \\ 
GSD18 & 53.071292 & -27.705802 &      25.24 $ \pm $       0.04 &        861 $ \pm $         66 &      -2.36 $ \pm $       0.11 &       7.85 $ \pm $       0.17 \\ 
GSD19 & 53.181976 & -27.705038 &      25.71 $ \pm $       0.06 &       1002 $ \pm $        245 &      -2.18 $ \pm $       0.41 &       7.72 $ \pm $       0.23 \\ 
GSD20 & 53.140815 & -27.692390 &      26.23 $ \pm $       0.10 &        496 $ \pm $         82 &      -1.83 $ \pm $       0.26 &       7.88 $ \pm $       0.21 \\ 
GSD21 & 53.100936 & -27.676704 &      24.76 $ \pm $       0.10 &        935 $ \pm $        139 &      -0.95 $ \pm $       0.28 &       8.54 $ \pm $       0.21 \\ 
GSD22 & 53.118450 & -27.819919 &      26.76 $ \pm $       0.13 &        870 $ \pm $        198 &      -1.88 $ \pm $       0.36 &       7.41 $ \pm $       0.24 \\ 
GSD23 & 53.077606 & -27.812795 &      26.81 $ \pm $       0.17 &       1512 $ \pm $        338 &      -2.22 $ \pm $       0.28 &       7.30 $ \pm $       0.25 \\ 
GSD24 & 53.132972 & -27.740102 &      27.77 $ \pm $       0.30 &        698 $ \pm $        318 &      -2.23 $ \pm $       0.47 &       6.95 $ \pm $       0.38 \\ 
GSD25 & 53.084388 & -27.727920 &      27.29 $ \pm $       0.15 &        562 $ \pm $        164 &      -2.39 $ \pm $       0.43 &       7.18 $ \pm $       0.27 \\ 
GSD26 & 53.141502 & -27.724880 &      26.65 $ \pm $       0.10 &        650 $ \pm $        106 &      -2.10 $ \pm $       0.25 &       7.47 $ \pm $       0.21 \\ 
GSD27 & 53.112579 & -27.707090 &      26.91 $ \pm $       0.13 &        954 $ \pm $        262 &      -2.40 $ \pm $       0.42 &       7.15 $ \pm $       0.26 \\ 
GSD28 & 53.046119 & -27.705604 &      27.13 $ \pm $       0.16 &       1009 $ \pm $        293 &      -2.14 $ \pm $       0.45 &       7.16 $ \pm $       0.28 \\ 
GSD29 & 53.139953 & -27.675138 &      27.79 $ \pm $       0.21 &       1314 $ \pm $        557 &      -2.12 $ \pm $       0.74 &       6.93 $ \pm $       0.37 \\ 
UDS1 & 34.275299 & -5.274496 &      25.38 $ \pm $       0.09 &        576 $ \pm $         90 &      -1.33 $ \pm $       0.18 &       8.31 $ \pm $       0.20 \\ 
UDS2 & 34.440769 & -5.262566 &      25.74 $ \pm $       0.09 &       1081 $ \pm $        147 &      -1.41 $ \pm $       0.17 &       7.99 $ \pm $       0.20 \\ 
UDS3 & 34.482173 & -5.261399 &      25.28 $ \pm $       0.08 &        507 $ \pm $         95 &      -1.88 $ \pm $       0.24 &       8.23 $ \pm $       0.20 \\ 
UDS4 & 34.268657 & -5.260064 &      25.44 $ \pm $       0.10 &        614 $ \pm $         83 &      -1.53 $ \pm $       0.14 &       8.18 $ \pm $       0.19 \\ 
UDS5 & 34.426483 & -5.255770 &      25.69 $ \pm $       0.11 &        701 $ \pm $         95 &      -1.66 $ \pm $       0.10 &       7.99 $ \pm $       0.20 \\ 
UDS6 & 34.428569 & -5.255318 &      25.10 $ \pm $       0.07 &        731 $ \pm $         86 &      -2.12 $ \pm $       0.13 &       8.04 $ \pm $       0.18 \\ 
UDS7 & 34.325676 & -5.251743 &      24.32 $ \pm $       0.04 &        656 $ \pm $         43 &      -1.59 $ \pm $       0.05 &       8.58 $ \pm $       0.17 \\ 
UDS8 & 34.314014 & -5.251047 &      26.44 $ \pm $       0.17 &        728 $ \pm $        153 &      -1.39 $ \pm $       0.20 &       7.77 $ \pm $       0.24 \\ 
UDS9 & 34.382587 & -5.244620 &      25.94 $ \pm $       0.09 &        478 $ \pm $         64 &      -1.82 $ \pm $       0.14 &       8.03 $ \pm $       0.19 \\ 
UDS10 & 34.263534 & -5.239433 &      25.48 $ \pm $       0.07 &        541 $ \pm $         64 &      -1.84 $ \pm $       0.14 &       8.13 $ \pm $       0.18 \\ 
UDS11 & 34.311279 & -5.238957 &      26.36 $ \pm $       0.10 &        735 $ \pm $         94 &      -2.43 $ \pm $       0.12 &       7.42 $ \pm $       0.19 \\ 
UDS12 & 34.473888 & -5.234232 &      24.15 $ \pm $       0.03 &        713 $ \pm $         42 &      -1.72 $ \pm $       0.05 &       8.57 $ \pm $       0.16 \\ 
UDS13 & 34.318141 & -5.232299 &      25.35 $ \pm $       0.07 &        716 $ \pm $         68 &      -2.12 $ \pm $       0.08 &       7.95 $ \pm $       0.18 \\ 
UDS14 & 34.481567 & -5.222499 &      25.69 $ \pm $       0.11 &        602 $ \pm $         96 &      -2.30 $ \pm $       0.15 &       7.81 $ \pm $       0.20 \\ 
UDS15 & 34.371166 & -5.214803 &      25.45 $ \pm $       0.09 &        843 $ \pm $        111 &      -1.35 $ \pm $       0.15 &       8.14 $ \pm $       0.19 \\ 
UDS16 & 34.482921 & -5.214187 &      25.39 $ \pm $       0.08 &        662 $ \pm $         87 &      -1.80 $ \pm $       0.15 &       8.07 $ \pm $       0.19 \\ 
UDS17 & 34.247516 & -5.205330 &      25.95 $ \pm $       0.10 &        469 $ \pm $         63 &      -2.22 $ \pm $       0.13 &       7.89 $ \pm $       0.19 \\ 
UDS18 & 34.315448 & -5.200902 &      25.83 $ \pm $       0.13 &        739 $ \pm $        136 &      -1.87 $ \pm $       0.17 &       7.84 $ \pm $       0.21 \\ 
UDS19 & 34.298866 & -5.191800 &      26.25 $ \pm $       0.15 &        543 $ \pm $        106 &      -1.94 $ \pm $       0.17 &       7.79 $ \pm $       0.23 \\ 
UDS20 & 34.232082 & -5.190388 &      25.16 $ \pm $       0.06 &        648 $ \pm $         55 &      -2.20 $ \pm $       0.06 &       8.03 $ \pm $       0.17 \\ 
UDS21 & 34.308940 & -5.190090 &      25.15 $ \pm $       0.07 &        609 $ \pm $         66 &      -2.28 $ \pm $       0.11 &       8.03 $ \pm $       0.18 \\ 
UDS22 & 34.416740 & -5.180443 &      26.34 $ \pm $       0.19 &       1070 $ \pm $        307 &      -1.72 $ \pm $       0.31 &       7.64 $ \pm $       0.28 \\ 
UDS23 & 34.387023 & -5.177240 &      25.59 $ \pm $       0.07 &        591 $ \pm $         74 &      -2.15 $ \pm $       0.14 &       7.91 $ \pm $       0.18 \\ 
UDS24 & 34.252845 & -5.176362 &      26.02 $ \pm $       0.13 &        779 $ \pm $        140 &      -1.40 $ \pm $       0.19 &       7.91 $ \pm $       0.22 \\ 
UDS25 & 34.402145 & -5.175352 &      24.47 $ \pm $       0.05 &        507 $ \pm $         39 &      -2.11 $ \pm $       0.06 &       8.48 $ \pm $       0.17 \\ 
UDS26 & 34.459190 & -5.174448 &      25.34 $ \pm $       0.08 &        552 $ \pm $         59 &      -2.29 $ \pm $       0.09 &       8.01 $ \pm $       0.18 \\ 
UDS27 & 34.284194 & -5.164084 &      26.40 $ \pm $       0.17 &        576 $ \pm $        139 &      -1.05 $ \pm $       0.30 &       8.01 $ \pm $       0.26 \\ 
UDS28 & 34.500236 & -5.155595 &      25.89 $ \pm $       0.14 &        519 $ \pm $        100 &      -1.77 $ \pm $       0.21 &       8.01 $ \pm $       0.22 \\ 
UDS29 & 34.263267 & -5.152174 &      26.35 $ \pm $       0.13 &       1003 $ \pm $        150 &      -1.90 $ \pm $       0.11 &       7.57 $ \pm $       0.21 \\ 
UDS30 & 34.477771 & -5.147521 &      25.66 $ \pm $       0.14 &        533 $ \pm $        103 &      -0.65 $ \pm $       0.21 &       8.50 $ \pm $       0.22 \\ 
UDS31 & 34.296325 & -5.144416 &      25.95 $ \pm $       0.12 &        546 $ \pm $         88 &      -2.17 $ \pm $       0.14 &       7.81 $ \pm $       0.20 \\ 
UDS32 & 34.419242 & -5.142892 &      25.83 $ \pm $       0.12 &        721 $ \pm $        138 &      -1.69 $ \pm $       0.21 &       7.91 $ \pm $       0.22 \\ 
UDS33 & 34.246810 & -5.139120 &      26.06 $ \pm $       0.11 &        553 $ \pm $         85 &      -1.54 $ \pm $       0.19 &       7.99 $ \pm $       0.20 \\ 
UDS34 & 34.371933 & -5.137272 &      24.83 $ \pm $       0.05 &        586 $ \pm $         50 &      -2.73 $ \pm $       0.08 &       8.01 $ \pm $       0.17 \\ 
UDS35 & 34.314693 & -5.133675 &      26.42 $ \pm $       0.21 &       1125 $ \pm $        315 &      -1.20 $ \pm $       0.28 &       7.80 $ \pm $       0.28 \\ 
UDS36 & 34.261756 & -5.134672 &      25.32 $ \pm $       0.08 &        658 $ \pm $         76 &      -2.16 $ \pm $       0.12 &       7.97 $ \pm $       0.18 \\ 
UDS37 & 34.380569 & -5.268105 &      26.54 $ \pm $       0.22 &        832 $ \pm $        249 &      -1.34 $ \pm $       0.28 &       7.71 $ \pm $       0.29 \\ 
UDS38 & 34.441444 & -5.215963 &      26.60 $ \pm $       0.19 &        912 $ \pm $        208 &      -2.14 $ \pm $       0.14 &       7.38 $ \pm $       0.25 \\ 
UDS39 & 34.334854 & -5.177163 &      26.99 $ \pm $       0.18 &        594 $ \pm $        145 &      -2.39 $ \pm $       0.24 &       7.26 $ \pm $       0.26 \\ 
UDS40 & 34.438095 & -5.160070 &      26.86 $ \pm $       0.17 &        677 $ \pm $        157 &      -2.29 $ \pm $       0.21 &       7.30 $ \pm $       0.25 \\ 
\enddata
\tablecomments{EELG2011: identification number prefixed by the
  respective field acronyms; RA/DEC: coordinates from the CANDELS
  catalogs; $H$: $H$-band AB magnitude from the CANDELS catalog;
  $EW_{\rm{[OIII]},5007}$: rest-frame equivalent width inferred from the
  $I$, $J$, and $H$ broad-band photometry (see text for details);
  $\beta_{2500}$: $F_{\lambda}$ continuum slope at rest-frame
  2500$\rm{\AA}$ inferred from a linear fit to the $B$, $V$, and $I$
  broad-band photometry; $\log{(M)}$: stellar mass inferred from
  Starburst99 \citep{leitherer99}, as described in the text.}
\end{deluxetable*}

\section{Discussion}

\subsection{Comparison with Other Samples}

Galaxies with similar properties have previously been identified
through broad-band photometry at $z<0.4$ in the Sloan Digital Sky
Survey \citep{cardamone09}, and have been shown by \citet{amorin10}
and \citet {izotov11} to constitute the most strongly star-forming and
most metal poor tail of the well-known class of blue compact dwarf
galaxies \citep[e.g.,][]{sargent70, thuan81, griffith11}, which have
very low metallicities and extremely high, spatially concentrated
star-formation activity \citep{guzman98, overzier08}.

\citet{cowie11} \citep[also see][]{scarlata09}) recently studied the
Ly$\alpha$ properties of high-EW H$\alpha$ emitters, providing a
direct connection between higher-redshift searches of Ly$\alpha$
\citep[e.g.,][]{ouchi08, hu10}, and find Ly$\alpha$ EWs ranging from
20$\rm{\AA}$ to 200$\rm{\AA}$.  Combining this with the findings of
\citet{nilsson11}, who show that Ly$\alpha$ emitters at $z\sim 2$ are
objects with a very wide range in properties, it is clear that from
Ly$\alpha$ emitters one cannot derive a complete description of star
formation in low-mass galaxies.  On the other hand, Ly$\alpha$
emitters at higher redshifts ($z>3$) appear to be young, with small
stellar masses \citep[e.g.,][]{finkelstein09}, similar to the emission
line galaxies studied here.

Narrow-band surveys identified galaxies with strong [OIII] and
H$\alpha$ emission lines with $\rm{EW}\sim 100-1000\rm{\AA}$ at
redshifts $z=0.3-1$ \citep[e.g.,][]{kakazu07}, demonstrated to be
young and metal poor \citep{hu09}.  Most notably, \citet{atek10}
pointed out the existence of a class of emission line galaxies at
$z\sim 1.5$ with $\rm{EW}> 1000\rm{\AA}$ that would most likely be
included in our sample as well.  However, so far, their nature has not
been described and their cosmological relevance in the context of
galaxy formation has remained unclear.  Therefore, let us now put
these starbursting dwarf galaxies in a cosmological context.

\subsection{Cosmological Context: Implications for the Formation of
  Dwarf Galaxies}

Our sample with redshifts $1.6 < z < 1.8$ consists of 69 low-mass
($\sim 10^8~\msol$), young ($\sim 0.5-4 \times 10^7~\rm{yr}$), extreme
starbursting, presumably metal-poor galaxies.  Their co-moving number
density\footnote{the values for our two widely separate fields, UDS
  and GSD, differ by only 12\%} is $3.7\times10^{-4}~\rm{Mpc}^{-3}$,
two orders of magnitude higher than that of nearby galaxies with
similar EWs \citep{cardamone09}.  The individual star formation rates
and the number density combine into $1.7\times10^{-3} ~\msol
~\rm{yr}^{-1} ~\rm{Mpc}^{-3}$.  This is a 2\% contribution to the
total star-formation rate density at $z\sim 1.7$ occurring in galaxies
that contribute perhaps $\sim 0.1\%$ to the total stellar mass density
at that epoch \citep[e.g.,][]{karim11}.

Placing these detections in the context of the burst fraction among
equally massive dwarf galaxies is difficult, however, since we cannot
currently constrain their number density; galaxies without starbursts
have older luminosity weighted ages and, as a consequence, are at
least 2 magnitudes fainter than the starbursting galaxies.  With
marginal detections, even in the latest WFC3 data, and no obvious
spectral features, their redshifts cannot easily be estimated.  Model
predictions differ strongly from observational measurements of the
galaxy stellar mass function \citep{guo11} and cannot be used to
constrain the burst fraction among the population of low-mass
galaxies.

Nevertheless, we can still gauge the importance of the observed burst
for the formation of low-mass galaxies by making the reasonable (and
testable) assumption that the observed bursts occur with the same
frequency at all epochs $1\lesssim z \lesssim 3.5$ -- a period of
$\sim$4 Gyr during which the cosmic star formation history peaked and
after which the number density of starbursts declines as mentioned
above.  The basic indication that such bursts are important is that
the number of stars produced in such bursts over a period of several
Gyr is comparable to the number stars in present-day dwarf galaxies.
Let us use this consideration to construct a simple toy model that
relates the observations presented here to the mass function of
present-day low-mass galaxies.  \citet{guo11} use the data from
\citet{baldry08} for the galaxies with masses down to $10^7~\msol$,
and their mass function can be represented by a simple power law for
all galaxies with masses $<10^{10}~\msol$, that is, well below the
knee of the \citet{schechter76} function:
%\begin{equation} \frac{\phi(M_*)}{\rm{Mpc}^3~\rm{dex(M)}} \sim 0.043
%  \Big(\frac{M_*}{10^8~\msol}\Big)^{-0.37}.
%\end{equation}
\begin{equation} \phi(M_*) \sim 0.043
  \Big(\frac{M_*}{10^8~\msol}\Big)^{-0.37},
\end{equation}
%\begin{equation} \log{\phi(M_*)}d\log{M_*} \sim -1.37 - 0.37
%  \log \Big(\frac{M_*}{10^8~\msol}\Big).
%\end{equation}
in units of $\rm{Mpc}^{-3}~\rm{d}\log(M_*)^{-1}$.

Let us then express the stellar mass of the present-day descendants of
the observed starbursting galaxies at $z\sim 1.7$ in terms of the
following star formation history:
\begin{equation} M_{\rm{desc}} = \frac{M_{\rm{burst}} \times
    N_{\rm{burst}}}{f_{\rm{burst}}},
\end{equation}
where $M_{\rm{burst}}\sim 2 \times 15~\rm{Myr} \times
5~\msol~\rm{yr}^{-1} = 1.5\times 10^8~\msol$ is the total stellar mass
produced in a single starburst (the factor 2 is included to convert
the observed burst age to its total duration), $N_{\rm{burst}}$ is the
number of bursts that occurs in each galaxy over the $\sim$4 Gyr
period between $z=1$ and $z=3.5$, and $f_{\rm{burst}}$ is the fraction
of the total stellar mass that is produced in such bursts -- the rest
is assumed to form in smaller bursts and/or a more quiescent mode of
star formation.

Now we can predict the number of present-day descendants by dividing
the number of observed bursts at $z\sim 1.7$ by the duty cycle, that
is, the fraction of the time that a bursts is occurring over the
period of time that they \textit{can} occur (here assumed to be
$1\lesssim z \lesssim 3.5$, or 4 Gyr).  This duty cycle is the ratio
of the duration of a single burst ($2\times 15$~Myr) and this 4 Gyr
period, multiplied by the number of bursts per galaxy
$N_{\rm{burst}}$. Thus we have

\begin{equation} \phi(M_{\rm{desc}}) \sim
  \frac{\phi_{\rm{burst}}}{N_{\rm{burst}}} \times
  \frac{4~\rm{Gyr}}{30~\rm{Myr}},
\end{equation}
where $\phi_{\rm{burst}}$ is the co-moving number density of the
bursts observed at $z\sim 1.7$
($3.7\times10^{-4}~\rm{Mpc}^{-3}~\rm{d}\log(M_*)^{-1}$, where we use
the observed span in stellar masses of an order of magnitude -- see
Figure 6 -- to introduce this differential unit).

According to Eqs.~2 and 4 the mass of the descendant, $M_{\rm{desc}}$,
is uniquely determined by the fraction of stars formed in bursts,
$f_{\rm{burst}}$.  Equating $\phi(M_{\rm{desc}})$ (Eq.~4) and
$\phi(M_*)$ (Eq.~2), and substituting for $N_{\rm{burst}}$ (Eq.~3) we
obtain, after rearranging terms:
\begin{equation} \frac{M_{\rm{desc}}}{10^8~\msol} \sim
  2.4f_{\rm{burst}}^{-1.6}.
\end{equation}

Thus, if $f_{\rm{burst}}$ is close to unity, that is, almost all stars
are formed in bursts, then we infer that each galaxy must undergo one
or two bursts on average and that, therefore, $M_{\rm{desc}} \sim 1-2
M_{\rm{burst}}$.  It is perhaps more realistic to adopt a smaller
value for $f_{\rm{burst}}$.  If $f_{\rm{burst}}\sim 0.5$, we find that
two or three bursts must occur in each galaxy, producing descendants
with masses $\sim 10^{9}~\msol$.  The latter would imply a growth in
stellar mass between $z\sim 1.7$ and the present by at least a factor
3 given the mass constraints on the underlying populations of the
observed starburst galaxies.  Galaxy formation models suggest that the
typical growth is indeed a factor of 3 or 4, but we should bear in
mind that these models do not reproduce the observed co-moving number
density evolution of low-mass galaxies with redshift
\citep[e.g.,][]{guo11}.  Therefore, these predictions should be
treated with care.

Choosing $f_{\rm{burst}}$ very low ($\lesssim 0.1$) implies a very
large growth in mass, with high-mass descendants ($>10^{10}~\msol$).
Such growth by, say, more than an order of magnitude in mass seems
unlikely in the context of current galaxy formation models. In
particular, models are better observationally constrained for these
higher masses and the prediction is that more massive galaxies
($10^{10-11}~\msol$) grow in mass by a factor 3 or 4, not by two
orders of magnitude, between $z\sim 1.7$ and the present
\citep[e.g.,][]{keres09}.  Given these constraints, the general
conclusion we can draw is that our observations suggest that many or
most stars in present-day dwarf galaxies (with masses $\lesssim
10^9~\msol$) have formed in a small number of starbursts at $z>1$.

The quantitative interpretation we offer here relies on our estimate
of the characteristic burst duration (30 Myr).  However, the
uncertainty in this assumed burst duration does not affect our
conclusion that intense starbursts at high redshift constitute an
important phase in the mass buildup low-mass present-day galaxies.  

If the bursts last significantly longer, which we deem unlikely, then
the total mass formed in a burst will be larger than assumed above,
while the frequency of the bursts will remain the same since our
selection method is only sensitive to young bursts.  As a result, the
fraction of stars formed in bursts, $f_{\rm{burst}}$ will be larger
than what we derived above.

Significantly shorter bursts (say, $\sim 5-10$~Myr, as allowed by the
SB99 model -- see Sec.~4.2) are also unlikely, as the bursts are
spatially extended, and the implied crossing time in these systems is
about 30 Myr.  Nonetheless, in the case that burst times are short,
the burst frequency must be higher to account for their observed
number (Eq.~4), but the total mass produced in each burst remains the
same (see Sec.~4.2).  As a result, the number of bursts per galaxy
($N_{\rm{burst}}$) and, therefore, the fraction of stars formed in
bursts ($f_{\rm{burst}}$) increase.

The main assumptions in our interpretation, then, are 1) that the
descendants of the observed galaxies remain low-mass galaxies up to
the present day ($\lesssim 10^9~\msol$ in stars), and 2) that the
observed bursts do not only occur at $z=1.6-1.8$, but are equally
frequent over a much broader redshift range ($1\lesssim z \lesssim
3.5$). The first assumption is supported by the understanding that
more massive galaxies are not expected to grow in stellar mass by
several orders of magnitude, as would be required if the descendants
of the observed starbursting galaxies are much more massive.  The
second assumption is straightforward to test observationally by
searches for similar objects over a wider redshift range.

\subsection{Energy Budget and Core Formation}

The most remarkable property of these galaxies are their growth rates
(specific star formation rates) of $20-200~\rm{Gyr}^{-1}$.  This is
far outside the realm of normal, more massive star forming galaxies,
which typically have $\sim 1~\rm{Gyr}^{-1}$ \citep[e.g.,][]{karim11}.

The amount of energy deposited into the interstellar medium through
winds and supernovae ($10^{56-57}$~erg) exceeds the binding energy by
an order of magnitude, even if the total mass is larger than the
stellar mass by many factors.  This implies that the gas that is
fueling the starburst may be in the process of being blown out,
resulting in the end of the starburst phase.  Whether the gas will be
blown out of the halo depends on the halo mass, which is currently
unconstrained.  If the gas does not leave the halo it may eventually
once more cool and sink to the center of the potential well and
trigger another starburst.  Then, the observed bursts could be part of
a semi-periodic cycle of star-formation activity, which is seen in
simulations of low-mass galaxies \citep[e.g.,][]{stinson07}.  Episodic
star formation would also alleviate the difficulty of explaining the
sudden occurence of such a large starburst in such small galaxies.  If
such starbursts are not cyclical, then another possible explanation,
usually offered to explain the lack of star formation at early times
in even less massive systems, is that star formation had been
suppressed at earlier epochs as a result of UV background radiation
\citep[e.g.,][]{babul92, babul96}.

An intriguing possibility is that if most of the gas is expelled from
the central region, then the stars themselves could become unbound as
well, dissolving the entire galaxy.  Displacement of large amounts of
material leads to changes in the potential, and if this happens at
very short time scales, then even the dark matter profile can be
altered.  Outflows have long been argued to play a role in producing
cored density profiles in low-mass galaxies
\citep[e.g.,][]{navarro96}.  The time scales, masses, and
star-formation rates we derived above for the galaxies in our sample
roughly meet the requirements for such a process to be efficient.
Recently, \citet{pontzen11} showed that episodic star formation, even
with less intense bursts, can produce the same effect.  What the
previous and subsequent evolution of the systems we observed is
remains to be seen, but the energy balance and the possibility that
such bursts are reoccurring make our observations consistent with this
general picture; indeed outflows may be responsible for the formation
of flattened dark matter profiles in low mass galaxies.  These
speculative scenarios can be tested further with observational
constraints on the gas masses and hydrodynamical modeling of these
systems.

\section{Summary}

Our discovery of an abundant galaxy population at $z\sim 1.7$ with
extremely high emission line equivalent widths implies that many
high-redshift, low-mass galaxies form many of their stars in extreme
starbursts.  We propose that we have observed an important formation
mode for dwarf galaxies: a small number of strong starbursts that
occur at early epochs ($z>1$) each form $\sim 10^{8}~\msol$ in stars
in a very short time span ($\sim 30$~Myr) to build up the bulk of the
stellar components of present-day dwarf galaxies.  This is in
quantitative agreement with 'archaeological' studies of present-day
dwarf galaxies, which have shown that their star formation histories
are burst-like and that the ages of their stellar populations suggest
formation redshifts $z>1$ \citep[e.g.,][]{weisz11}.  Under the
reasonable assumption based on $\Lambda$CDM predictions for galaxy
grwoth that the observed galaxies grow in mass by less than an order
of magnitude up to the present day, our observations provide direct
evidence for such an early formation epoch and, in particular, that
short-lived bursts contribute much or even the majority of star
formation in dwarf galaxies.

\textit{Facilities:} HST(ACS,WFC3), VLT(VIMOS), Spitzer(IRAC)

\acknowledgements{A.v.d.W.~thanks the following people for useful
  feedback and stimulating discussions: Greg Stinson, Andrea Maccio,
  Brent Groves, Dan Weisz, Joe Hennawi, Kate Rubin, Sharon Meidt, and
  Marijn Franx.  S.M.F., J.R.T., D.C.K., D.D.K., K.L., and
  E.G.M. acknowledge funding through HST GO-12060 and NSF AST
  08-08133.}

%\bibliography{/Users/vdwel/papers/bibtex/mypapers}{}
\bibliographystyle{apj}

\end{document}